# Micromachined force scale for optical power measurement by radiation pressure sensing

Ivan Ryger, Alexandra B. Artusio-Glimpse, Paul Williams, Nathan Tomlin, Michelle Stephens, Kyle Rogers, Matthew Spidell, and John Lehman

National Institute of Standards and Technology, 325 Broadway, Boulder, CO 80305

## Abstract
We introduce a micromachined force scale for laser power measurement by means of radiation pressure sensing. With this technique, the measured laser light is not absorbed and can be utilized while being measured. We employ silicon micromachining technology to construct a miniature force scale, opening the potential to its use for fast in-line laser process monitoring. Here we describe the mechanical sensing principle and conversion to an electrical signal. We further outline an electrostatic force substitution process for nulling of the radiation pressure force on the sensor mirror. Finally, we look at the performance of a proof-of-concept device in open-loop operation (without the nulling electrostatic force) subjected to a modulated laser at 250 W and find its response time is less than 20 ms with noise floor dominated by electronics at 2.5 W/√Hz.

## Introduction

Laser based manufacturing is a rapidly growing industry that is increasing demand for accurately controlled lasers. One of the key factors influencing the microstructure of welds, and thus their quality, is the amount of energy delivered to the melt spot. To address fast monitoring of laser power throughout a weld process, laser manufacturers typically include a beam splitter with high splitting ratio in the optical path of the beam to pick off a small portion of the laser energy for fast measurement with a photodectector. However, large beam splitting ratios, factors on the order of $10^5$-$10^6$, are needed to prevent damage to the monitoring detector because the power level of these lasers is on the order of kilowatts. As such, errors in the absolute splitting ratio of parts-in-a-million, common from heating of the optic, can easily lead to measurement errors on the order of magnitude of the measurand [1]. In contrast, the most accurate laser power measurement technique is the calorimetric measurement principle [2, 3]. This is by its nature a very slow process involving total absorption of light, thus precluding monitoring of power in-process. For this reason, a paradigm shift is required to enable in-line laser-based process characterization with high accuracy.

Radiation pressure measurements are the most viable means for in situ, high accuracy laser power measurements. The pressure induced by optical radiation was first described by J. C. Maxwell [4] and later physically demonstrated by Nichols and Hull [5]. Since then, several authors have developed sensitive force scales for precise measurement of photon-induced pressure [6-10], but these complex and expensive instruments have required special laboratory conditions, prohibiting them from widespread implementation.

The portable laser power meter employing a commercially available force scale for radiation pressure measurements was first introduced by P. Williams, et al. [11]. This power meter brought about a considerable shift in how laser power measurement is approached. Classically, accurate high-power laser measurements and use of the laser beam were two mutually exclusive operations. Authors of the radiation pressure power meter (RPPM) demonstrated an accurate and traceable measurement with a relatively portable device and reported a rigorous study of the measurement uncertainties [12]. The lower measurement limit of the RPPM is limited by acoustic vibrations, and scale sensitivity and response time. Although vibrations can be reduced to a reasonable extent in a laboratory environment, they are harder to control in a manufacturing environment. For this reason, we conceived of a radiation-pressure-based laser power meter employing micromachined dual spring detector concept, where two identical springs are used in a tandem configuration that mitigates environmental vibration signals as well as errors due to changes in the sensor tilt.

The system centers around two micromachined silicon springs that allow us to reduce the total dimensions and inertial mass of the sensor, improving both the sensitivity and speed. Much in the way of optical, mechanical, electrical, and control system development is required in creating this new, compact radiation pressure power meter. In this article, we constrain ourselves to describe the basic geometry of the device and detail the sensing electronics used to convert the radiation pressure force of an incident laser to an electrical signal. We outline the process for calibrating this electrical signal both in an open-loop configuration where the sensor response is mapped prior to use and in a closed-loop configuration where the radiation pressure force is balanced with a known electrostatic force. We end the paper with a brief example of the data acquired from laser injection tests on a sample device and a summary of the device performance and error sources.

## Micromachined Scale and its Sensing Principle

The dual spring sensor is depicted in Figure 1(a). This silicon micromachined force scale consists of a mirror-coated silicon disk *b* attached to a supporting silicon ring *a* through thin spring legs *c* in the shape of an Archimedean spiral. The disk deflection is sensed by a capacitance change in a parallel plate capacitor through a pair of conductive electrodes *d* on the backside of each disk. Electrical connection *e* to the sensing electronics is provided through metallized extended pads *f*. Both capacitor elements (chips) are clamped together by a machined plastic mount with a predefined spacing (42.5 µm) given by a plastic insulator in the shape of an annular ring.

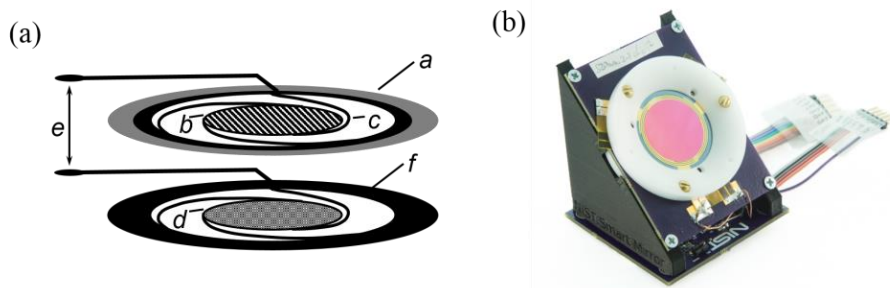

*Figure 1. (a) Schematic representation of the sensor mechanical system. (b) Photograph of a fabricated sensor.*

A simplified "single-spring" version of Figure 1(a) consists of the upper mirror/electrode/spring assembly as shown, but with the bottom assembly having electrode *d* fixed in position with respect to the outer annulus f. The single-spring implementation precludes vibrational and gravitational compensation but is a simpler design used for preliminary device testing. A photograph of a fully clamped and mounted sensor is shown in Figure 1(b). The sensing element consists of a silicon disk of 20 mm diameter attached to a silicon annulus through three narrow spiral legs (width 265 µm, thickness 380 µm, length 45 mm).

Fabrication of the sensing element starts with a 76.2 mm (3 inch) double-side polished silicon wafer with 160 nm layer of thermally grown oxide. The highly reflective Bragg mirror (optimized for peak reflectivity of 1.07 µm light at 45° incidence - measured to be 0.9999±0.0001) is deposited on one side of the wafer and etched to cover only the central disk of each chip. Using a rigorous coupled wave analysis to predict the mirror properties in conjunction with spectrophotometric reflectance and calorimetric absorptance measurements, the mirror has an optical bandwidth of 1035-1125 nm for a reflectance greater than 0.999. According to predictions, angular variation of 5° affects the reflectance at the fourth decimal place (ΔR=±0.0002). The capacitor electrode (Ti/Au) is deposited on the opposite side of the wafer by e-beam evaporation. In preliminary experiments, we observed a strong thermal dependence of the sensor capacitance; therefore, we added a Ti/Au strain balancing film to the spring legs on the mirror side to compensate the thermal expansion coefficient (CTE) mismatch between the metal along the back of the spring legs and silicon. Lastly, the disk reflector and spiral legs are patterned by deep reactive ion etching of the substrate. Fabricated sensor disks are then sandwiched around the thin annulus of polyimide film and held together by a machined clamp made from polyoxymethylene resin.

Impinging optical photons exert a weak force on the surface of the moveable mirror, pushing one plate of the sensing capacitor toward the other. This causes a fractional change to the sensor capacitance that can be measured electronically. The mirror deflection *Δh* can be calculated as

$$\Delta h = \frac{2rP}{ck}\cos(\theta) \quad (1)$$

where *P* is the optical power, *c* the speed of light, *k* the spring constant, and θ the angle of incidence of the beam on the mirror surface. The reflectivity-dependent constant $r = R + (1-R)\alpha/2$, , accounts for the fact that an absorbed photon transfers all its momentum, while a reflected photon transfers twice its momentum, where R is the reflectivity of the Bragg mirror and α is the fraction of non-reflected light that is absorbed by the mirror.

In the arrangement of a dual spring sensor, the bottom conductive plane *d* (Figure 1(a)) is a micromachined silicon chip identical to the top one. In the presence of common-mode mechanical vibrations, both top and bottom electrode will move in synchrony, keeping the spacing between them constant, thus suppressing the contribution of the vibrational noise and tilt-dependent gravity projection. A detailed investigation of the complete mechanical behavior of this dual spring system is beyond the scope of the present manuscript. In this paper, we will assume the dual spring system is ideal, thus returns no signal from vibration or rotation with respect to gravity.

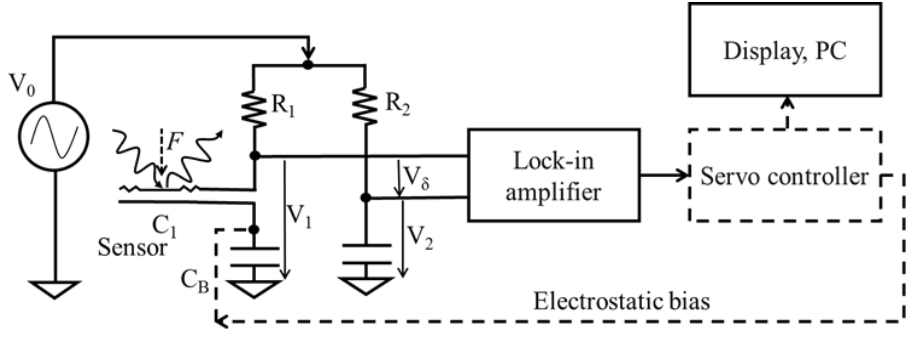

*Figure 2. Simplified block diagram of a capacitive force scale with electrical substitution; showing the optical force applied to variable capacitor C1. The force-dependent AC bridge signal is measured by lock-in amplifier. A servo controller could be employed for "closed-loop" control of the electrode spacing of C1. A large-value capacitor CB serves to decouple the DC electrostatic bias path from the AC bridge electronics readout.*

### Electronic Readout

A fractional change of the sensor capacitance (due to forcing from laser pressure on the mirror) is converted to electrical voltage in a capacitive bridge driven by a sinusoidal voltage with amplitude $V_0$ and frequency ω (Figure 2) [13,14]. In this null-indicator configuration, a low-noise high-gain amplifier can be employed. This is advantageous when the sensor is used with a closed feedback loop. The output signal from the bridge is amplified and synchronously demodulated in a digital lock-in amplifier. This approach allows us to suppress the presence of $1/f$ preamplifier noise [15] and low frequency interference. The output voltage phasor is decomposed into two orthogonal components (in-phase and quadrature or real and imaginary) that can be analyzed separately. Separation of the phasor in this way suppresses electromagnetic crosstalk from neighboring signal traces that is phase-shifted with respect to the sensor signal. In open-loop operation, the signal stops here and is recorded by a connecting PC.

In the closed-loop configuration, a servo controller pre-deflects the sensor plate to a defined bias point by an electrostatic force [17,18]. When light impinges on the sensor surface, the plate spacing decreases. To compensate this reduced spacing, the controller then reduces the electrostatic force, returning the spring to its initial position. Because the membrane is effectively held in a fixed position, the sensor spring constant does not need to be known to determine the amount of applied force. Moreover, the conversion between measured force and electrostatic bias is given by the simple relationship

$$\Delta F_{ES} = -F_{PH} = \frac{\epsilon_0 A}{2h_p^2}\left(V_{ES(0)}^2 - V_{ES(1)}^2\right) \quad (2)$$

where $\Delta F_{ES}$ is the change in electrostatic force, $F_{PH}$ the photon force, $\varepsilon_0$ the permittivity of air, $A$ the capacitor electrode area, $h_p$ the plate spacing set point, and $V_{ES(0)}$ and $V_{ES(1)}$ are the applied electrostatic bias before and after application of photon force, respectively.

Closed-loop operation of the radiation pressure sensor is ultimately of interest to us as it eliminates steady state deflection of the sensor mirror (resulting is beam steering) and relieves the need to accurately map out the sensor responsivity to laser power. However, since a detailed discussion of the required feedback controller is beyond the scope of this article, in the following parts of this paper we focus on the sensor operation in the open-loop configuration where output from the lock-in amplifier is directly sent to the PC or display (Figure 2).

## Sensitivity estimation

Based on a linearized model of the system, we evaluate the system responsivity *S(P)* by applying the chain rule of derivatives

$$S(P) = \frac{dV_\delta}{dC_1}\frac{dC_1}{dh}\frac{dh}{dP} \quad (3)$$

where *dh/dP* represents the spring deflection upon application of the optical power, obtained by differentiating (1), where $\Delta h = h_0 - h$ is the deflection, $h_0$ is the initial capacitor plate spacing and *h* is the deflected plate spacing. The sensitivity of capacitance to plate deflection is $dC_1/dh = -\varepsilon_0 A/h^2$. We neglect the effects of fringing fields because the ratio between electrode diameter and plate separation is about 500, large enough to ignore edge effects.

The first term, $dV_\delta/dC_1$, in (3) represents the sensitivity of the bridge output difference voltage to the change of capacitance. Both arms of the capacitive bridge form a low-pass electrical filter for the drive signal $V_0$. We express the total output voltage of one bridge arm as

$$V_1(\omega) = V_0(\omega)(1 + j\omega\tau)^{-1} \quad (4)$$

where $\tau_1 = R_1 C_1$. Therefore, the bridge transfer function is dependent on selection of the excitation frequency $\omega = 2\pi f_0$ and component parameters $R_1$ and $C_1$. Since the blocking capacitor $C_B$ is large (~100 nF) compared to the active sensing capacitor, we can neglect it in our calculations. A similar transfer function can be written for the reference bridge arm. Then the bridge difference voltage will be $V_\delta = V_1 - V_2$ (Figure 2). Differentiating with respect to $\tau_1$, we get a sensitivity of the complex voltage $V_\delta$ as a function of frequency and bridge parameters. The second bridge arm will not contribute to (3) since its output is constant. Therefore, we denote the sensitivity $dV_\delta/dC_1$ as a product of three factors: $\omega$, $R_1$, and a dimensionless complex function *H*.

$$\frac{dV_\delta(\omega)}{dC_1} = \frac{R_1 dV_\delta(\omega)}{d\tau_1} = \omega[H_R(\Theta) + jH_I(\Theta)]R_1 \quad (5)$$

The complex function of normalized parameter $\Theta = \omega\tau_1$ is split into its real part

$$H_R(\Theta) = -\Theta(1 + \Theta^2)^{-2} \quad (6)$$

and imaginary part

$$H_I(\Theta) = -(1 - \Theta^2)(1 + \Theta^2)^{-2} \quad (7)$$

A plot of both parts of the dimensionless function is shown in Figure 3 as a function of the product $\Theta = \omega\tau_1$. For small values of $\Theta$, the imaginary function $H_I$ is slowly varying and close to -1. This implies a very linear response of the imaginary part of the output voltage $V_\delta$ with respect to the change of parameter $\tau_1$. As the value of $\Theta$ rises, the absolute value of $H_I$ decreases, and after $\Theta$ passes the critical value of $\Theta \sim 1$, the derivative changes sign. A designer should avoid this point as it brings about ambiguity in the sensor state detection. The real function $H_R$ has a local maximum at $\Theta \sim 0.6$ and does not change sign, but brings non-linearities into the measurement. In our first prototype reported further below, the parameter $\Theta$ was fixed at an operating point of 1.4 rad.

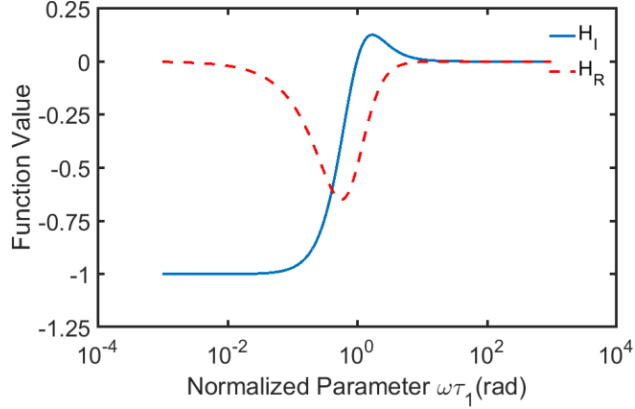

*Figure 3. Plot of two orthogonal components H_R (in-phase) and H_I (quadrature). of the dimensionless sensitivity function.*

### Sensor noise

Dominant sources of electrical noise influencing the system signal to noise ratio are the bridge resistors and the instrumentation amplifier (Figure 4). Because these noise sources are uncorrelated, the energy carried by them adds in quadrature [15] and resulting total noise voltage $e_n^{tot}$ at the input of the low noise preamplifier is

$$e_n^{TOT} = \left[4k_BT(R_1 + R_2) + e_{n,A}^2 + i_{n,A}^2(R_1 + R_2)^2\right]^{1/2} \qquad (8)$$

where the term $k_BT$ represents the Boltzmann's thermal energy, and $e_{n,A}$ and $i_{n,A}$ are the preamplifier voltage noise and current noise, respectively. Combining the total noise voltage with the previously calculated value of the sensor responsivity, we obtain the estimated noise equivalent power as $NEP = e_n^{tot}/S(P)$.

The effect of acoustic noise is not considered as this is an environment-dependent quantity. In a coming contribution, we will analyze acoustic and other environmental noise sources to a greater extent.

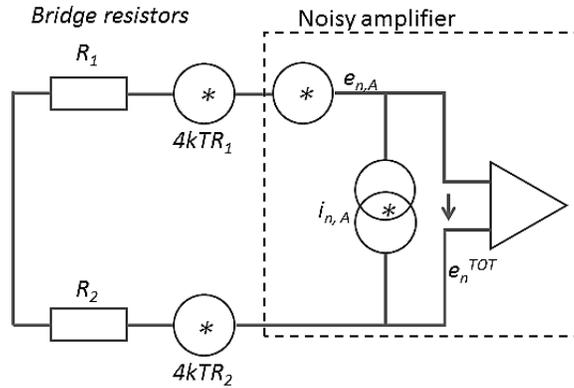

*Figure 4. Simplified noise model of the bridge electronics including the contribution of resistor and amplifier noise.*

### Open-loop calibration

Our electro-mechanical system is highly non-linear in the open-loop configuration; therefore, a careful characterization of the electro-mechanical system is required to determine the initial rest state of the system to properly estimate its responsivity. Fortunately, the sensor can be opto-electrically characterized while packaged, simplifying the calibration process. In this section, a smaller single spring device (mirror diameter D= 10 mm, spring constant k=42 N/m) is used in these plots for demonstration of the calibration procedure.

## Static characterization

When the system is operated in open loop, the two most critical parameters affecting the mechanical system responsivity are the spring constant $k$ and capacitor initial electrode spacing $h_0$. For this reason, we developed a measurement method that allows us to determine these two parameters simultaneously. This is done by applying an electrostatic force between the sensor capacitor plates and measuring the distance between capacitor electrodes $h$ with an interferometer. The balance between applied electrostatic force and spring restoring force is expressed by

$$\epsilon_0 A V^2 / (2h^2) = k(h_0 - h) \tag{9}$$

With the interferometer, we measure the change in electrode spacing $\Delta h = h_0 - h$ instead of the absolute spacing $h$; therefore, we rewrite equation (9) in the following polynomial form

$$\epsilon_0 A V^2 / (2k) = \Delta h^3 - 2h_0 \Delta h^2 + h_0^2 \Delta h \tag{10}$$

By fitting the inverse of the measured dependence $V^2=f(\Delta h)$ to a cubic polynomial (10)

$$\frac{\epsilon_0 A V^2}{2} = a_3 \Delta h^3 + a_2 \Delta h + a_1 \Delta h \tag{11}$$

and matching appropriate coefficients we get

$$a_3 = k, \quad -a_2/(2a_3) = h_0, \quad -2a_1/a_2 = h_0 \tag{12}$$

The interferometer (wavelength $\lambda_i$ = 632.8 nm) is setup in a cat's eye [19] arrangement (Figure 5) for measurement of deflection $\Delta h$. This provides a separation between forward and return path beams. A translation stage is used for proper placement of an achromatic doublet plano-convex lens one focal length from the sensor and centered so that forward and return (reflected) path beams pass through the same radial location of the lens.

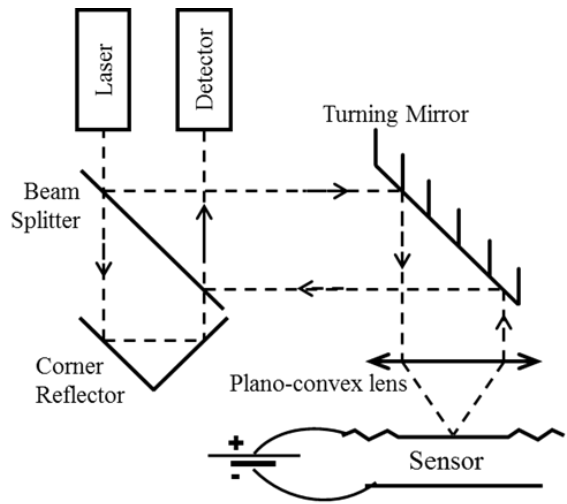

Figure 5. Interferometric deflection measurement setup in cat's eye arrangement for electro-mechanical calibration.

An example of measured deflection with applied electrostatic voltage together with the polynomial approximation is shown in Figure 6. From this measurement, we obtained a fit of the spring constant $k$ = 45.8 N.m$^{-1}$ and initial electrode spacing of $h_0$ = 38.6 µm. This is in good agreement with simulation, which predicts a spring constant of 42 N.m$^{-1}$. The spacing agrees well with the measured thickness of the spacer (between sensor plates); which was measured to range from 30 to 50 µm.

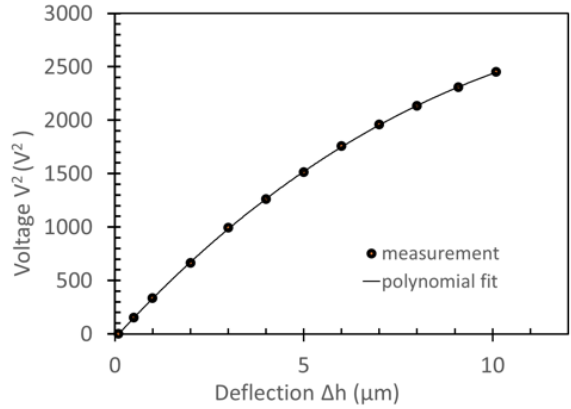

*Figure 6. Inverse dependence of spring deflection with applied electrostatic voltage for spring constant determination.*

To verify the robustness of this algorithm, we simulate noisy voltage data over the range $V_{MIN}$ to $V_{MAX}$ with Gaussian distributed noise defined by standard deviation $\sigma_v$ with a number of measured points $N$. The spring constant used for the simulation was $k$ = 42 N.m$^{-1}$ and electrode spacing was $h_0$ = 45 µm. Results of the algorithm response to simulated noisy data are summarized in Table 1. The accuracy of the fitting curve can be easily checked instantaneously by comparing the values of height $h_0$ obtained from ratios - $a_2/(2a_3)$ and $-2a_1/a_2$, which should agree.

*Table 1. Sensitivity of the fitting algorithm to noise in data*

| N | $\sigma_V$ (V) | $V_{MIN}..V_{MAX}$ (V) | $k$ (N/m) | $h_0$ (µm) $2a_1/a_2$ | $h_0$ (µm) $a_2/(2a_3)$ | $\delta k$ (%) | max {$\delta h_0$} (%) |
|---|---|---|---|---|---|---|---|
| 11 | 0 | 0..50 | 42 | 44.9 | 45 | 0.0 | 0.2 |
| 101 | 0.1 | 0..50 | 41.6 | 45.1 | 45.2 | -1.0 | 0.4 |
| 101 | 1 | 0..50 | 31.4 | 47.4 | 56.8 | -25 | 26 |
| 101 | 0.01 | 20..30 | 129.7 | 10.4 | 61.2 | 209 | 123 |
| 101 | 0.01 | 0..10 | 3900 | -0.3 | 70.9 | 9386 | 101 |

From Table 1 we see that for accurate determination of fitting polynomial coefficients it is desirable that the measurement interval covers a range where the second and third derivatives of the function (11) are pronounced. Because this algorithm depends on coefficients of high order polynomial derivatives, it is vulnerable to the presence of noise in the data. In the case of noisy data, it is helpful to increase the measurement voltage span and number of measurement points.

### Capacitance Mapping

An accurate knowledge of the parasitic capacitance and initial capacitor plate spacing is important from a calibration point of view if the sensor is being operated in open-loop. Knowledge of these quantities is also needed for the development of a feedback controller design. The total sensor capacitance consists of two components: the active capacitance $C_s(\Delta h)$ (dependent on deflection) and the parasitic parallel capacitance $C_p$ of the sensor support and connecting leads (13). In our case, we neglected the contribution of the fringing field [20, 21] as it does not introduce a substantial deviation from this simple relationship:

$$C_1 = C_s(\Delta h) + C_p = \epsilon_0 A (h_0 - \Delta h)^{-1} + C_p \qquad (13)$$

Both capacitance quantities can be determined by measuring the total sensor capacitance ($C_1$) as a function of changing plate spacing. For this reason, a similar setup to the deflection measurement is prepared. While applying the electrostatic bias to induce deflection, the sensor capacitance is measured by a commercial LCR bridge with DC voltage bias option. A slightly different sensor (from a different batch) with a lower initial plate spacing (compared with sensor used in Figure 6) was used in the example measurement data plotted in Figure 7. The dependence between capacitor plate deflection and total sensor capacitance, along with the approximated hyperbolic dependence (13), is shown in Figure 7.

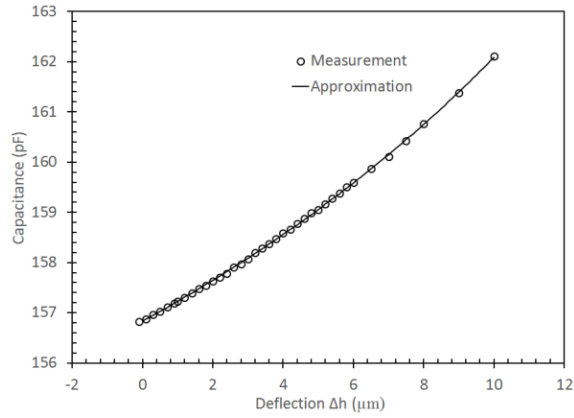

Figure 7. Dependence of sensor capacitance on moveable plate deflection ∆h for capacitance mapping.

Since in this configuration the mechanical system driven by electrostatic force is highly non-linear, there exists a region of unstable deflections where the attractive force diverges [20, 22]. Close to the boundary of this region, any small perturbation (vibrations or step voltage change) can trigger the pull-in effect [22]. To avoid pull-in, the range of deflections has to be limited to 1/3 of the initial electrode spacing.

Using a Levenberg-Marquardt [23] gradient optimization algorithm, the measured data are fit to equation (13). This algorithm is advantageous because it provides fast and stable convergence in this case. From the data in Figure 7, we obtain a value for the parasitic capacitance $C_P$ = 140.2 pF and initial electrode spacing $h_0$ = 41.9 µm. We also tested a Taylor polynomial fit of this capacitance mapping measured data to obtain these the parasitic capacitance and initial plate spacing. However, the Taylor polynomial fit was found to be unreliable in the presence of noisy data. In contrast, the gradient optimization method is more consistent in the presence of noise.

### Dynamic characterization

The sensor is found to have a sharp resonance around 120.8 Hz. For this reason, for closed-loop operation, the electronic servo controller design must include a characterization of the spring resonance to suppress ringing. During the measurement, special attention must be paid to recover the resonance signal from wideband noise due to the high quality factor of the spring. In an experiment, the sensor is excited with a tunable sinusoidal source connected to the voice coil of a shaker providing acoustic excitation (Figure 8).

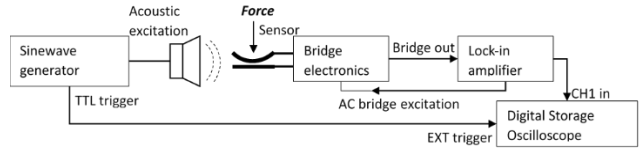

Figure 8. Acoustic sensor resonance measurement block diagram.

Output of the sensor signal is processed by a lock-in amplifier, with a low-pass cut-off frequency set to 1 kHz. This bandwidth is chosen to be beyond the maximum frequency of applied vibrations. Such a wide bandwidth, however, brings an excessive amount of white noise from the electronic frontend. For this reason, post-filtering is performed on the oscilloscope stage. The oscilloscope is triggered synchronously by a sinewave generator and waveform averaging is applied. This suppresses uncorrelated white noise, whereas the coherent sinewave response is amplified.

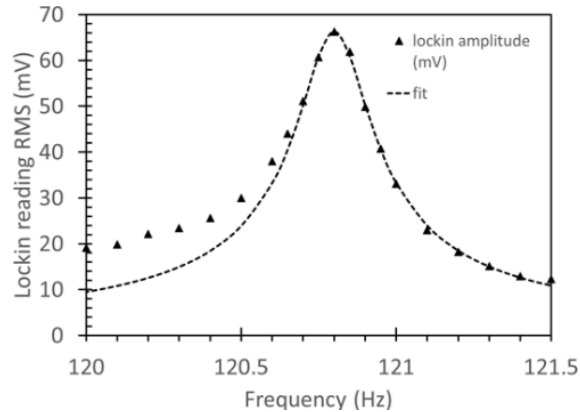

Figure 9. Resonance characteristics of a fabricated sensor with single-sided gold coating (CTE balance coating excluded).

In Figure 9, the measured spectrum is approximated by an analytical Lorenz curve [24] given by equation (14). From its parameters, we obtained the value of peak resonance frequency $f_0$=120.8 Hz and quality factor $Q$=520.

$$V(f) = V_{peak}/[1 + Q^2(f/f_0 - f_0/f)] \tag{14}$$

### Laser experiment

To demonstrate a radiation pressure signal and distinguish it from other phenomena, such as thermal or photoconductive effects, we performed the following tests. The mirror surface of the sensor is illuminated with a laser beam coming from a Yb- doped fiber laser (Figure 10). An adjustable iris aperture was used to prevent scattered light from illuminating the spring legs not covered by the Bragg reflector mirror. The beam reflected from the sensor's mirror surface was directed to a beam dump 1 m from the sensor. Some scattered light off the beam dump did return to illuminate the sensor, as seen with an IR-viewer. Due both to scattered light from the beam dump and to over filling of the sensor mirror by the wings of the multimode beam, large thermal drifting of the sensor was observed. To keep this drifting from saturating the sensor, we had to decrease the sinewave drive magnitude on the bridge circuit. This in turn yielded a decreased sensitivity.

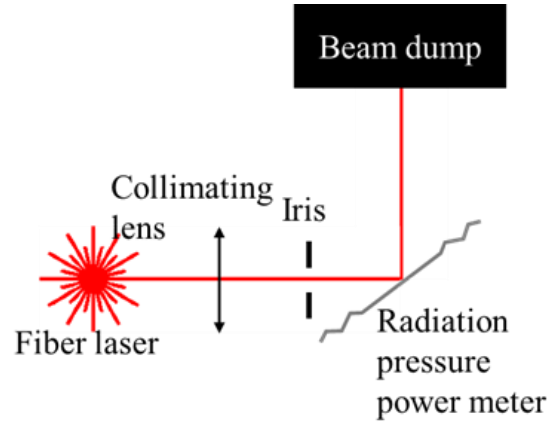

*Figure 10. Optical test setup for radiation pressure sensing.*

We used a dual spring configuration with two identical silicon chips, with the electrodes facing each other. The electrode sides of these chips were fully covered with gold leaving no exposed silicon in order to exclude the possibility of the photoconductive effect influencing the measured signal because the gold shields the sensing capacitor from any small change in the silicon conductance. Note, however, that this large area of gold increases the parasitic capacitance of the sensor, further reducing its sensitivity. A better way to shield against photoconductive effects is to make the springs out of an insulator rather than a semi-conductor like silicon. We are in the process of developing a fabrication technique to do just this in coming versions of this device.

The design parameters of this sensor follow: mirror diameter D=20 mm, capacitor plate spacing $h_0$=18 µm, spring stiffness k=80N/m, total sensor capacitance $C_1$=273 pF, bridge resistance $R_1$=41 kΩ, sinewave drive magnitude $V_0$=50 mV, and filter bandwidth 100 Hz. Thus, the predicted bridge sensitivity is 0.0157 µV/W. This multiplied by the preamplifier gain 1000× gives us the conversion between optical power and recorded electrical voltage. During the experiment, we chose to display the in-phase signal since it gave us stronger sensitivity to mirror deflection than the orthogonal component. The laser power incident on the sensor was 250 W modulated with 20% duty cycle at 5 Hz.

An example of transient data record from this sensor is shown in Figure 11, where baseline drifting of the signal is attributed to primarily heating of the sensor springs through absorption of the laser light both in the mirror and at the surface of the legs, where scattered light illuminates the area outside of the mirror. The fast pulses (when the beam is on) have negative amplitude, which corresponds to an increase in capacitance (electrodes are pushed towards each other), and agrees with sensitivity calculations. The slow thermal drift (time constant on the order of 1 minute) is the result of cantilever bending of spring legs as their temperature progressively rises due absorption of un-reflected light at the mirror and scattered light on the spring legs. Theoretically, a perfectly-matching CTE strain compensating layer should fully suppress the cantilever bending. In comparing springs with and without the strain-balancing metal layer, we observe fivefold suppression of the thermal drift. This corresponds to less than 1.5% thickness difference between the gold on the electrode side of the spring legs and the strain-balancing gold layer on the opposite side, and agrees with deposition accuracy.

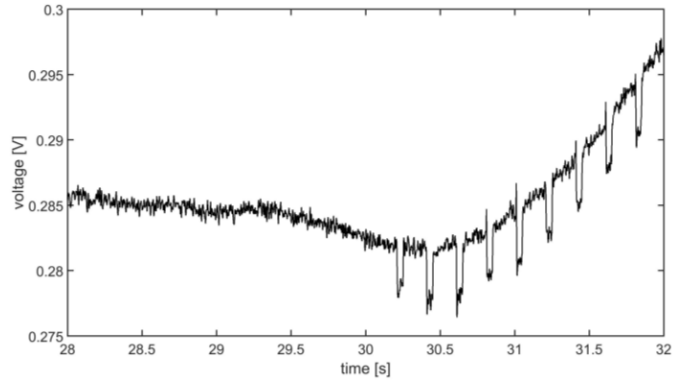

*Figure 11. Detailed view of sensor response before and after turning on modulated laser power showing slow drift in opposite to the radiation pressure signal.*

In Figure 12, the slow baseline drift in the signal from Figure 11 is subtracted. From this data, we measure an average signal of (3.71 ± 0.40) mV over the course of 10 pulses, 5% lower than the signal expected from sensitivity calculations (3.91 mV), The reason for this discrepancy is the accuracy of non-linear capacitive fit.

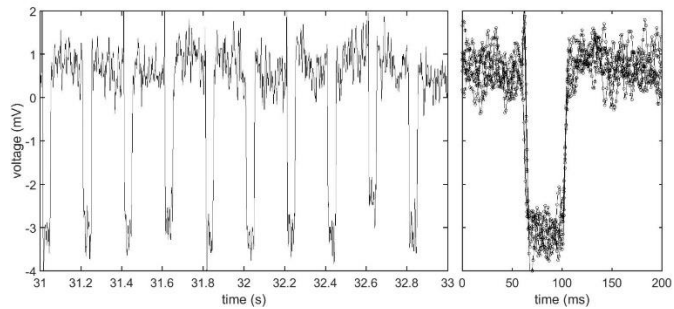

*Figure 12. (left) Detail of transient sensor response to pulsed laser power after subtraction of the baseline drift. (right) Overlay of 5 laser pulses showing the repeatability of the measurement within 10%, mostly caused by presence of noise. Laser beam of 250 W.*

The rise time of the signal is dominated by the sensor spring resonance and quality factor and measured to be 10-15 ms. The RMS value of signal noise is 0.36 mV integrated over 100 Hz filter bandwidth. This agrees well with our predicted noise (0.38 mV), corresponding to 2.5 W/√Hz NEP.

To convince ourselves that the fast signal corresponding to laser illumination of the sensor is indeed due to radiation pressure and not to other effects, we completed a series of tests and calculations to isolate thermal and photoconductive effects. To begin with, according to our calculations the temperature gradient across the spring leg cross-section (from the side of illumination to the side of electrodes) vanishes within 1 μs. This is much faster than our spring-mass time constant (~10 ms); therefore, fast transient effects due to heat absorption at the top of the spring legs cannot excite measurable deflection of springs. From four-wire measurements of the spring leg resistance with temperature using a spring with a mirror reflectance (R=0.999) worse than that used in the laser experiment, we find the temperature of the sensor follows a classic exponential heating curve with a time constant of 32 s and thermalizes 30°C above room temperature when continuously illuminated with a 500 W laser at 45° incidence. This trend agrees with the slow baseline drift highlighted in Figure 11, where, due to the higher reflectance, the time constant of drift is approximately 1 minute. We next measured the capacitance between the mirror side of a single spring and the electrode side of the same spring when illuminated by a laser directly on the legs or on the supporting silicon annulus. The intention was to exaggerate silicon absorption of photons that will generate electron-hole pairs that can modify the conductivity of the bulk silicon and effectively add to the parasitic capacitance of the sensor. We measured no capacitance change that depended on laser illumination (silicon absorption of photons). Entirely coating the electrode side of the spring with gold does shield the sensor from variation of silicon conductivity at the cost of statically increasing the parasitic capacitance. These tests (and calculation) support the argument that radiation pressure is being measured in the above and not thermal or photoconductive effects.

## Conclusions

In this contribution, we demonstrate a design process and characterization of a micromachined force scale for measurement of weak forces induced by photon pressure. Using a silicon micromachining technology, the size and inertial mass were substantially reduced making this sensor compact and faster than its bulk commercial scale-based predecessor [12]. We achieved 40-times reduction of the noise background and more than 250-times improvement in the measurement speed compared to [12]. These features make this device a strong candidate for fast and accurate in-line laser-based process control. Future work will be dedicated to the reduction of thermal drifting and parasitic capacitance for improved sensitivity and to the close-loop controller. The last will improve linearity and facilitate a more stable calibration process.

## Disclaimer

## References

[1] M. Spidell, H. Hadler, M. Stephens, P. Williams, and J. Lehman, "Geometric contributions to chopper wheel optical attenuation and uncertainty," *Metrologia* vol. 54, no. 4, pp. L19-L25, June 2017

[2] D. J. Livigni, C.L. Cromer, T.R. Scott, B. C. Johnson, Z. M. Zhang; "Thermal characterization of a cryogenic radiometer and comparison with a laser calorimeter," *Metrologia* vol. 35, no. 6, pp. 819-827, 1998

[3] E. D. West, W. E. Case, A. L. Rasmussen, and L. B. Schmidt, "A Reference Calorimeter for Laser Energy Measurements," *J. Res. Nat. Bur. Stand.* vol. 76A, no. 1, pp. 13-26, Jan.-Feb. 1972

[4] J. C. Maxwell, *A Treatise on Electricity and Magnetism*, 1st ed. (Oxford University, 1873)

[5] E. E. Nichols and G. F. Hull, "The pressure due to radiation," *Phys. Rev.* vol. 17, no. 5, pp. 315-351, Jun. 1903

[6] J. Cook, W. L. Flowers, and C. B. Arnold, "Measurement of Laser Output by Light Pressure," Proc. IRE vol. 50, pp.1693, May 1962.

[7] Y. P. Yuan, "A New Pulse Laser Energy Meter," *Rev. Sci. Instrum*. vol. 61 no. 6, pp. 1743–1746, Mar. 1990.

[8] V. Nesterov, M. Mueller, L. L. Frumin, and U. Brand, "A new facility to realize a nanonewton force standard based on electrostatic methods," *Metrologia*, vol. 46, no. 3, pp. 277–282, Apr. 2009

[9] K. Agatsuma, D. Friedrich, S. Ballmer, G. DeSalvo, S. Sakata, E. Nishida, and S. Kawamura, "Precise measurement of laser power using an optomechanical system," *Opt. Express* vol. 22, no. 2, pp. 2013–2030, Jan. 2014

[10] J. Stirling, F. G. Cervantes, J. R. Pratt, and G. A. Shaw, "A self-calibrating optomechanical force sensor with femtonewton resolution," *Appl. Phys. Lett.* vol. 105, no. 23, pp. 233109, Dec. 2014

[11] P. A. Williams, J. A. Hadler, R. Lee, F. C. Maring, and J. H. Lehman, "Use of radiation pressure for measurement of high-power laser emission," *Opt. Lett.* vol. 38, no. 20, pp 4248–4251, Oct. 2013

[12] P. Williams, J. A. Hadler, F. C. Maring, R. Lee, K. A. Rogers, B. J. Simonds, M. T. Spidell, A. D. Feldman, J. H. Lehman., "Portable, high-accuracy, non-absorbing laser power measurement at kilowatt levels by means of radiation pressure," *Opt. Express*, vol. 25, no. 4, pp. 4382-4392, Feb. 2017

[13] H. K. P. Neubert, "Capacitance transducers in a.c. bridge circuits," in "*Instrument Transducers- An introduction to their performance and design*," second edition, Oxford, Claredon Press, 1975 pp., 238-243

[14] V. Josselin, P. Toubul, R. Kielbasa, "Capacitive detection scheme for space accelerometers applications," *Sensor. Actuator.* vol. 78, no. 2-3, pp. 92–98, Dec. 1999

[15] C. D. Motchenbacher and J. A. Connely, "*Low Noise Electronic Design*," New York, Wiley, 1993, (pages 16-20, 38-41)

[16] I. Ryger, D. Harber, M. Stephens, M. White, N. Tomlin, M. Spidell, J. Lehman, "Noise characteristics of thermistors: Measurement methods and results of selected devices," *Rev. Sci. Instrum*. vol. 88, no. 024707, Feb. 2017; doi: 10.1063/1.4976029

[17] G. A. Shaw, J. Stirling, J. A. Kramar, A. Moses, P. Abbott, R. Steiner, A. Koffman, J. R. Pratt, Z. J. Kubarych, "Miligram mass metrology using an electrostatic force balance," *Metrologia* vol. 53, no. 5, pp. A86–A94, Sep. 2016, doi:10.1088/0026-1394/53/5/A86f

[18] L. Hugill, H. D. Valliant, "Limitations to the application of electrostatic feedback in gravity meters," *J. Geophys. Res.* vol. 91, B8., pp. 8387-8392, Jul. 1986

[19] F. E. Peña-Arellano, C. C. Speake, "Mirror tilt immunity interferometry with a cat's eye retroreflector," *Appl. Optics* vol. 50, no. 7, pp. 981-991, Feb. 2011

[20] M. Bao, "4.1.3. Fringe effects; 4.2.1. (1) Voltage Driving and the Pull-in Effect," in "*Analysis and Design Principles of MEMS Devices*," San Diego, Elsevier, 2005, pp. 178-186

[21] K. P. P. Pillai, "Fringing field of finite parallel-plate capacitors," Proc. IEE vol. 117, no. 6, pp. 1201-1204, Jun. 1970

[22] V. Kaajakari, *MEMS tutorial: Pull-in voltage in electrostatic microactuators*, [online] available http://www.kaajakari.net/~ville/research/tutorials/pull_in_tutorial.pdf

[23] J. C. T. Kelley, "Iterative Methods for Optimization," (the Society for Industrial and Applied Mathematics, 1999) [online] available http://www.siam.org/books/textbooks/fr18_book.pdf

[24] L. Scharf, "A First Course in Electrical and Computer Engineering," Rice University Houston, Texas, [online] available http://cnx.org/content/col10685/1.2